\documentclass[submission,copyright,creativecommons]{eptcs}
\providecommand{\event}{QPL 2014} % Name of the event you are submitting to
\usepackage{breakurl}             % Not needed if you use pdflatex only.

\usepackage{amsmath,amssymb,amsthm,txfonts,braket}

  \newtheorem{Def}{Definition}[section]
  \newtheorem{thm}[Def]{Theorem}
  \newtheorem{prop}[Def]{Proposition}
  \newtheorem{prob}[Def]{Problem}
  
  \newtheorem{cor}[Def]{Corollary}
  \newtheorem*{cl}{Claim}
  \newtheorem{conj}[Def]{Conjecture}

\theoremstyle{definition}
  \newtheorem{ex}[Def]{Examples}
  \newtheorem*{rem}{Remark}
  \newtheorem*{prf}{Proof}
  \newtheorem*{prf11}{Proof of theorem 2.1}
  \newtheorem*{prf46}{Proof of proposition 3.10}

\title{On G\'acs' quantum algorithmic entropy
\thanks{Most part of this research was carried out without knowing about
Tadaki's work [15]. Quite recently, in March 2014, Prof. Tadaki
draw our attention to his work and we noticed that there are substantial
overlaps between Tadaki's work and ours. As far as we know, his
work seems to be the first one providing an extension of Gacs' work
to the infinite-dimensional setting. In the present paper, we tried to
reflect Tadaki's results as possible as we can. Interested readers
should consult Tadaki's article [15]. We are very grateful to
Prof. Tadaki for his advices.}
}

\author{Toru Takisaka
\institute{Research Institute for Mathematical Sciences\\
Kyoto University\\
Kyoto, Japan}
\email{takisaka@kurims.kyoto-u.ac.jp}
}
\def\titlerunning{On G\'acs' quantum algorithmic entropy}
\def\authorrunning{Toru Takisaka}
\begin{document}
\maketitle

\begin{abstract}
We define an infinite dimensional modification of lower-semicomputability of density operators by G\'acs with an attempt to fix some problem in the paper. Our attempt is partly achieved by showing the existence of universal operator under some additional assumption. It is left as a future task to eliminate this assumption. We also see some properties and examples which stimulate further research. In particular, we show that universal operator has certain nontrivial form if it exists.
\end{abstract}

\section{Preliminaries}

Kolmogorov complexity is the notion of actual information content of finite string in computational point of view. This notion has been proposed by Kolmogorov - Solomonoff - Chaitin in 1960s and used in various areas as a basic tool to represent descriptive complexity. On the other hand, Since Shor's algorithm [1] has been discovered, the research on quantum information has made a great progress and produced various proposals on application to quantum information technology. 

Quantum Kolmogorov complexity is one of these branches appeared in early 2000s. Several different definitions are proposed so far [2-4], and some applications to quantum information are recently emerging [5-6]. However, it seems that there is very little progress in this area despite a decade has passed since these suggestions have been made, and a number of elementary facts are still not investigated.

In particular, relationships between them are not clarified. In classical domain, there are several definitions of descriptive complexity and some of them are known as equivalent notions (Levin's coding theorem). This theorem, in some sense, guarantees that these notions are reliable. 

It naturally leads us to the following question: can we find any good relationship between these quantum complexities? In particular, if it turns out that some of them are equivalent, it would be helpful to make these notions more reliable and more applicable to other research subject such like quantum information theory.

We particularly have interest on those by Berthiaume et al. [2] and G\'acs [3] since they are the quantum extension of plain Kolmogorov complexity and universal semimeasure, respectively. Levin's coding theorem claims that prefix Kolmogorov complexity and universal semimeasure are equivalent, so they are expected to be “nearly equivalent”.

For Berthiaume's definition, there are several results about fundamental facts such like its invariance and relation between classical complexity [7-9]. As compared to this, there are not so much subsequent research of G\'acs' approach, so we mainly treat his definition.

In [3], the quantum analogue of lower-semicomputable semimeasure which is named {\it lower-semicomputable semi-density matrix} is introduced. In the paper, though, proofs of two crucial theorems have some flaw.
% (the notion of lower-semicomputability of operators, an infinite dimensional modification of G\'acs' definition, will be introduced in section 3)
\begin{conj}
There is a lower-semicomputable semi-density matrix $\mu$ dominating all other such matrices in the sense that for every other such matrix $\rho$ there is a constant $c > 0$ with $\rho \leq c\mu$. 
\end{conj}
\begin{conj}
Let $\ket{1}$, $\ket{2}$, $\ldots$ be a computable orthogonal sequence of states. Also Let $\overline{H}$ and $\underline{H}$ be real-valued functions defined as
\[
 \overline{H}(\ket{\psi}) = -\braket{\psi|(\log \mu)\psi}, \mbox{    }  \underline{H}(\ket{\psi}) =-\log \braket{\psi|\mu\psi}.
\]
Then for $H = \overline{H}$ or $H = \underline{H}$ we have
\[
H(\ket{i}) = K(i) + O(1).
\]
Here, $K(i)$ is the prefix Kolmogorov complexity of $i$.
\end{conj}
The former is indispensable to define quantum algorithmic entropy, and the latter is expected to be true when we wish to compare G\'acs' quantum algorithmic entropy and the qubit complexity defined by Berthiaume et al [2].

In this paper, we introduce an infinite dimensional modification of G\'acs' definition to fix these problems. Our attempt is partly achieved by showing the existence of universal operator under some additional assumption. This is an analogous approach to the one of Tadaki [15], in which the notion of {\it lower-computable semi-POVM} is introduced, and it is shown that a universal semi-POVM does exist. Still, it seems that this assumption should be derived from our definition itself, so checking whether it is possible or not is our future task. It turns out that, in our modification, if we assume the existence of universal operator then Conjecture 1.2 is also true. We also see some properties and examples which stimulate further research.

Contents of this paper are as follows:
in section 2, we recall some classical notions of descriptive complexity for preparation. 
In section 3, we propose an infinite dimensional modification of lower-semicomputable semi-density matrix, which is defined by G\'acs to define his quantum algorithmic entropy. We prove some of their properties, and consider the problem about the existence of universal operator. 

We assume the readers are familiar with the basic ideas and technics of quantum information theory. The most famous textbook of this area would be Nielsen and Chuang [17], but we also suggest Heinosaari and Ziman [18] as an introduction, which is fairly readable and includes knowledge for infinite dimensional cases. For more exhaustive learning of functional analysis, see Conway [19].

\section{Classical notions of descriptive complexity}

In this section, we review two classical notions about descriptive complexity which are equivalent in some sense. Proof of any theorem in this section can be found in [12].

%2.1
\subsection{Kolmogorov complexity}

{\it(Plain) Kolmogorov complexity}  $ C_M (w) $  of finite binary string $ w $ with respect to a Turing machine $ M $ is the length of a shortest program which makes $ M $ output $ w $:

\[
   C_M (w)  = {\rm min}  \set{l(v)|M(v)=w}.
\] 

$M$ is called a {\it reference machine}. In many cases, some optimal universal Turing machine $M_0$ is employed as a fixed reference machine and $C(w)\coloneqq C_{M_0}(w)$ is just called Kolmogorov complexity of $w$. 
Here, we say $M_0$ is {\it optimal} if for any Turing machine $M$ there exists $c_M>0$ such that

\[
 C_{M_0}(w) \leq C_{M}(w) + c_M.
\]

$A \subset \set{0,1}^*$ is a {\it prefix set}  if for any two disjoint elements $w,v\in A,$ $w$ is not a prefix of $v$, and vice versa: that is, $w \neq vu$ and $v \neq wu$ for any $u \in \set{0,1}^*$. We call a Turing machine $T$ {\it Prefix Turing machine} if ${\rm dom}T$ is a prefix set. We can enumerate all prefix Turing machines effectively, and there exists an optimal universal prefix Turing machine. For detail, see [12]. We fix some optimal universal prefix Turing machine $M_1$ and call $K(w) \coloneqq C_{M_1}(w)$ {\it prefix Kolmogorov complexity} of $w$.

%2.2
\subsection{Lower-semicomputable semimeasure}

A nonnegative real function $ f(w) $ on strings is called {\it a semimeasure} if $ \sum_{w} f(w) \leq 1$, and a {\it measure} if the sum is 1. {\it f} is {\it lower-semicomputable} if there is a computable function $ \tilde {f} : \set{0,1}^* \times \mathbb{N} \to \mathbb{Q} $ such that $ \tilde {f} (w,k) \leq \tilde {f} (w,k+1) $ for every $w \in \set{0,1}^*$, $k \in \mathbb{N}$, and $ \tilde{f} (w,k)  \xrightarrow{k \to \infty} f(w) $ for every $w$. We call $\tilde f$ a {\it lower-approximation} of $f$ (we use this notation for convenience, but probably this function does not have any widely accepted name).

\begin{thm}
We can enumerate all lower-semicomputable semimeasures effectively. Namely, there exists $\tilde m : \set{0,1}^* \times \mathbb{N}^2 \to \mathbb{Q}$ which satisfies following two conditions:
\begin{enumerate}
  \item for any $n \in \mathbb{N}$, $\tilde m (-,-,n)$ is a lower-approximation of some lower-semicomputable semimeasure;
  \item for given lower-semicomputable semimeasure  $m'$, there is $n \in \mathbb{N}$ such that $\tilde m (-,-,n)$ is a lower-approximation of $m'$.
\end{enumerate}
\end{thm}
It is well known that there exists a universal semimeasure in the following sense.

\begin{thm}

There is a semicomputable semimeasure $ {\rm \bf m}$ with the property that for any other semicomputable semimeasure $m'$ there is a constant $ c >0 $ such that for all w we have $c m'(w) \leq  {\rm \bf m}(w)$.

\end{thm}

\begin{prf}

We can easily show that

\[
 {\rm \bf m}(w) \coloneqq \sum_{n=1}^\infty 2^{-n}m_n(w)
\]
is a universal semimeasure, where $\set{m_n}_{n=1}^\infty$ is an effective enumeration of all lower-semicomputable semimeasures. \qed

\end{prf}

We conclude this section with a theorem due to Levin. It indicates that the notion of universal semimeasure is somewhat equivalent to that of Kolmogorov complexity.

\begin{thm}[Levin's coding theorem]

$K(w) = - \log {\rm \bf m}(w)+O(1)$.

\end{thm}

\section{Quantization of lower-semicomputable semimeasure}

In this section, we define an infinite dimensional modification of lower-semicomputable semi-density matrix defined by G\'acs [3], and see some properties, examples, and problems.

%3.1
\subsection{Definition and some properties}

%G\'acs [3] introduced quantum extension of lower-semicomputable semimeasure. He defined it as a sequence of matrices on finite dimensional Hilbelt space, but there are some flaws in his paper. Some of them can be fixed by modifying his definition to an infinite dimensional version, so we mainly discuss about this modified one.
As a quantum analogue of the set of all binary strings, we introduce the space of indeterminate-length qubit strings, $\mathcal{H} \coloneqq \bigoplus _{n=0}^\infty (\mathbb{C}^2)^{\otimes n}$. We assume an orthonormal basis $\set{\ket{0},\ket{1}}$ is given for each qubit space $\mathbb{C}^2$, so $\mathcal{H}$ has an orthonormal basis  $\set{\ket{w}}_{w \in \set{0,1}^*}$, where $\ket{w} = \ket{a_1} \otimes \cdots \otimes \ket{a_n}$ for $w = a_1 \cdots a_n$. We call it the {\it computational basis} of $\mathcal{H}$. Notice that the computational basis is indispensable to consider descriptive complexity of qubit strings, just as in classical domain we need to work on $\set{0,1}^*$, not $\omega$.

Let $\mathcal{B(H)}$ be the set of all bounded operator on $\mathcal{H}$, and $\mathcal{L(H)}$ be the set of all bounded hermitian operator on $\mathcal{H}$. 
We also write $\mathbb{C}_q \coloneqq \set{x+yi|x,y \in \mathbb{Q}}$.

\begin{Def}

$\rho \in \mathcal{L(H)}$ is called a {\rm semi-density operator} if $\rho \geq 0$ and ${\rm Tr} \rho \leq 1$. Let $\tilde {\mathcal{S}}(\mathcal{H})$ be the set of all semi-density operators on $\mathcal{H}$.  %使わないかもしれない($\tilde {\mathcal{S}}(\mathcal{H})$)

$\rho \in \mathcal{L(H)}$ is {\rm lower-semicomputable (upper-semicomputable)} if there is a computable function $\psi : \mathbb{N} \times \set{0,1}^* \times \set{0,1}^* \to  \mathbb{C}_q $ such that the sequence $\set{\rho_n}_{n=1}^\infty \subset \mathcal{L(H)}$ defined by 

\[
\braket{w|\rho_n v} \coloneqq \psi(n,w,v) 
\]
satisfies $\rho_n \leq \rho_{n+1}$ ($\rho_n \geq \rho_{n+1}$) and $\rho_n \xrightarrow{n \to \infty} \rho$ in WOT (i.e. $\braket{\psi|\rho_n \psi} \to \braket{\psi|\rho \psi}$ for any $\ket{\psi} \in \mathcal{H}$. WOT is an abbreviation of {\rm weak operator topology}). We call $\psi$ a {\rm lower- (upper-) approximation} of $\rho$.

$\rho \in \mathcal{L(H)}$ is {\rm computable} if there is a computable function $\psi : \mathbb{N} \times \set{0,1}^* \times \set{0,1}^* \to \mathbb{C}_q $ which defines $\set{\rho_n}_{n=1}^\infty \subset \mathcal{L(H)}$ such that $\| \rho - \rho_n \|  < 2^{-n}$, in the same manner as above. We call $\psi$ an {\rm approximation} of $\rho$.

\end{Def}
 
Dimension of the string space $\mathcal{H}$ is almost the only difference between the definition by G\'acs and us. A mode of convergence of $\set{\rho_n}_{n=1}^\infty$ needs to be specified when we work on an infinite dimensional space, so we choose WOT, which is equivalent to the pointwise convergence of each matrix coefficient.
G\'acs allows a lower-approximation function to take an algebraic number, but we do not feel it necessary, so we only allow a complex-rational value.

\begin{rem}

Perhaps lower-semicomputability of operator can be defined for unbounded hermitian operator, but we will content ourselves with this definition in this paper. We mainly treat lower-semicomputable semi-density operators, which are automatically bounded. 

\end{rem}

It is equivalent to define the correspondence between $\psi$ and $\set{\rho_n}_{n=1}^\infty$ as

\[
 \psi(n,w,v) = 
　\begin{cases}
　　\braket{w + v | \rho (w + v)} & (w \leq v) \\
　　\braket{w + iv | \rho (w + iv)} & (w > v).
　\end{cases}
\]
In this definition, a lower approximation $\psi$ of any lower-semicomputable semi-density operator satisfies $\psi(n,w,v) \leq \psi(n + 1,w,v)$. Also notice that the converse is not true; there exists $\set{\rho_n}_{n=1}^\infty$ which is not increasing but corresponding $\psi$ is increasing with respect to $n$. In fact, the matrix

\[ \rho \coloneqq
\left(%
\begin{array}{cccc}
  3            & 2            & 0        & \ldots \\
  2            & 1            & 0        & \ldots \\
  0            & 0            & 0        & \ldots \\
  \vdots     &  \vdots    & \vdots & \ddots \\
\end{array}%
\right)
\]
is not positive, but $\braket{w + v | \rho (w + v)} \geq 0$ and $\braket{w + iv | \rho (w + iv)} \geq 0$ hold for every $w,v \in \set{0,1}^*$.

In contrast, our computability of operator is equivalent to that of its approximation function.
\begin{prop}

$\rho$ is computable if and only if $\psi(w,v) = \braket{w|\rho v}$ is computable (in the classical sense).

\end{prop}

\begin{prf}

Suppose $\rho \in \mathcal{L(H)}$ is computable and let $\set{\rho_n}_{n=1}^\infty$ be an approximation of $\rho$. Then Schwarz inequality tells us 
\[
|\braket{w|(\rho - \rho_n) v}| 
     \leq \| \rho - \rho_n \|
     < 2^{-n},
\]
which shows that $\braket{w|\rho_n v}$ is an $n$-digit approximation of $\braket{w|\rho v}$.

Conversely, suppose $\psi(w,v) = \braket{w|\rho v}$ is computable, i.e. there is a computable function $\tilde \psi : \set{0,1}^* \times \mathbb{N}^2 \to \mathbb{C}_q$ such that $|\psi(w,v) - \tilde \psi(w,v,n)| < 2^{-n}$ for any $w,v,n$. Then 
$\tilde \varphi (w,v,n) \coloneqq \tilde \psi (w,v, \lceil \frac{w+v+n}{2}+1 \rceil)$
is an approximation  of $\rho$. In fact, let $\set{\sigma_n}_{n=1}^\infty$ be the sequence of operators induced by $\tilde \varphi$. then

\[
  \| \rho - \sigma_n \| 
      \leq \| \rho - \sigma_n \|_{HS} 
      \leq \sum_{w,v}|\psi(w,v) - \tilde \varphi(w,v,n)|^2
      < 2^{-n}.
\]

Here, $\| \cdot \|_{HS}$ is the Hilbert-Schmidt norm

\[
\| \rho \|_{HS} = \sum_{w,v}|\braket{w|\rho v}|^2.
\]
\qed

\end{prf}

In the classical domain, a function is computable if and only if it is lower- and upper-semicomputable. The same thing can be said in our quantum modification.

\begin{prop}

$\rho$ is computable if and only if it is lower-semicomputable and upper-semicomputable.

\end{prop}

\begin{prf}
Let $\rho$ be lower- and upper-semicomputable. Also let $\set{\underline \rho_n}_{n=1}^\infty$ and $\set{\overline \rho_n}_{n=1}^\infty$ be an lower- and upper approximation of $\rho$, respectively. Then  $\psi(w,v) = \braket{w|\rho v}$ is computable since
\begin{eqnarray*}
|\braket{w|(\rho - \overline \rho_n)v}|
   & \leq & \frac{1}{4} \sum_{k=0}^3 |\braket{w+i^k v|(\rho - \overline \rho_n)(w+i^k v)}| \\
   & \leq & \frac{1}{4} \sum_{k=0}^3 |\braket{w+i^k v|(\underline \rho_n - \overline \rho_n)(w+i^k v)}| \\
   & \xrightarrow{n \to \infty} & 0
\end{eqnarray*}
holds, and we can compute the right side of inequality successively for all $n$. This means we can construct a computable function $f : \mathbb{N} \to \mathbb{N}$ which makes $\tilde \psi (n,w,v) \coloneqq \braket{w|\rho_{f(n)}v}$ an approximation of $\rho$. Notice that this proof is slightly different from classical one since lower- and upper-approximation of $\rho$ itself is not a one of $\psi$.

%%%%%%%%%%前のthmを経由しない証明%%%%%%%%%%%
%Let $\rho$ be lower- and upper-semicomputable, and $\underline{\psi}$ and $\overline{\psi}$ be a lower and upper approximation of $\rho$, respectively. define $\varphi$ as

%\[
%\varphi(i,j,n) \coloneqq \overline{\psi}(i,j,k(i,j,n)) - \underline{\psi}(i,j,k(i,j,n)),
%\]

%where $k(i,j,n)$ is an integer which satisfies $ \overline{\psi}(i,j,k(i,j,n)) - \underline{\psi}(i,j,k(i,j,n)) < 2^{-(\frac{i+j+n}{2}+1)}$. $\varphi$ is computable and the sequence $\set{\sigma_n}_{n=1}^\infty$ induced by $\varphi$ satisfies

%\[
%\| \rho - \sigma_n \| \leq \| \rho - \sigma_n \|_{HS} \leq \sum_{i,j}|\varphi(i,j,n)|^2 < 2^{-n}.
%\]

%Here, $\| \cdot \|_{HS}$ is Hilbert-Schmidt norm

%\[
%\| \rho \|_{HS} = \sum_{i,j}|\braket{i|\rho j}|^2.
%\]
%%%%%%%%%%前のthmを経由しない証明（終）%%%%%%%%%%%

Conversely, let $\rho$ be computable. Then we can obtain a lower-approximation $\set{\tilde \rho_n}_{n=1}^\infty$ of $\rho$ defining

\[
\tilde \rho_n \coloneqq \rho_n - 2^{-n+2} I.
\]

In fact, $\| \rho - \tilde \rho_n \| \to 0 $ so $\tilde \rho_n \to \rho$ in WOT. Using the inequality $\rho \leq \| \rho \| I$ it can be shown

\[
\rho_n - \rho_{n+1} < 2^{-n+1}I.
\]

Hence

\[
\tilde \rho_n - \tilde \rho_{n+1} = \rho_n - \rho_{n+1} - 2^{-n+1}I \leq 0.
\]

Obviously $\set{\tilde \rho_n}_{n=1}^\infty$ is induced by a computable function: let $\tilde \psi (w,v,n) \coloneqq \psi (w,v,n) - 2^{-n+2} \delta_{ij} $. Upper-semicomputability of $\rho$ can be shown in the same manner. \qed

\end{prf}

We say a sequence $\set{\ket{\psi_n}}_{n=1}^\infty$ of states is {\it uniformly computable} if
there is a recursive function $\tilde \psi: \mathbb{N}^2 \times \set{0,1}^* \to \mathbb{C}_q$ such that  
\[
|\braket{w|\psi_n} - \tilde \psi(k,n,w)| < 2^{-k}
\]
for every $k,n \in \mathbb{N}$ and $w \in \set{0,1}^*$. 

Let $m$ be a lower-semicomputable semimeasure, and $\set{\ket{\psi_n}}_{n=1}^\infty$ be  a uniformly computable sequence of states. If it holds $\braket{w|\psi_n}\braket{\psi_n|v} \in \mathbb{C}_q$ for every $w,v \in \set{0,1}^*$ and $n \in \mathbb{N}$, then obviously an operator $\sum_n m(n)\ket{\psi_n}\bra{\psi_n}$ is lower-semicomputable: in fact, $\set{\sum_n \tilde m(k,n)\ket{\psi_n}\bra{\psi_n}}_{k=1}^\infty$ is its lower-approximation. It turns out that we can discard the last assumption.

\begin{prop}
Let $\set{\ket{\psi_n}}_{n=1}^\infty$ be a uniformly computable sequence of states, and m be a lower-semicomputable semimeasure. Then $\rho \coloneqq \sum_n m(n)\ket{\psi_n}\bra{\psi_n}$ is a lower-semicomputable semi-density operator.
\end{prop}

\begin{prf}
For every $k,n \in \mathbb{N}$ and $w \in \set{0,1}^*$, let $\ket{\psi_{k,n}}$ be a vector (not necessarily a state) which is identified by an equation
\[
\braket{w|\psi_{k,n}} = \tilde \psi (k+w,n,w).
\]
Then it is routine to show that $\tilde \rho_k' \coloneqq \sum_n \tilde m(n) \ket{\psi_{k,n}}\bra{\psi_{k,n}}$ converges to $\rho$ in WOT (actually it converges in norm). We can also show that $\tilde \rho_k \coloneqq \tilde \rho_k' - 2^{-(k+1)}I$ forms a lower-approximation of $\rho$ in the same manner as proposition 3.3. \qed
\end{prf}

It is still open whether the converse is also true or not. Formally, can we find a uniformly computable sequence $\set{\ket{\psi_n}}_{n=1}^\infty$ of states and a lower-semicomputable semimeasure $m$ such that $\rho = \sum_n m(n)\ket{\psi_n}\bra{\psi_n}$ for any lower-semicomputable semi-density operator $\rho$? But at least, we expect that taking $\set{\ket{\psi_n}}_{n=1}^\infty$ as an orthonormal basis is {\it not} always possible, since otherwise there is no universal operator, as we see in proposition 3.12 and corollary 3.13.

We conclude this subsection with some examples which shows that some obvious property in classical domain fails to hold in our quantum version. In classical case, it is always possible to take a sequence of {\it positive} functions as a lower-approximation of semimeasure, since if $\psi$ is a lower-approximation of $m$ then so is $\varphi (x,k) \coloneqq {\rm max} \set{\psi(x,k),0}$. This is not always true in our quantum modification.

\begin{ex}[{[15]}]
There is a lower-semicomputable semi-density operator which cannot be approximated by any sequence of positive operators from below.
In fact, let $\rho$ be a rank-one projection of which nonzero eigenvector is $\frac{1}{2}\ket{\lambda} + \frac{\sqrt{3}}{2}\ket{0}$. Matrix representation of $\rho$ is

\[
\frac{1}{4}
\left(%
\begin{array}{cccc}
  1            & \sqrt{3}    & 0        & \ldots \\
  \sqrt{3}    & 3            & 0        & \ldots \\
  0            & 0            & 0        & \ldots \\
  \vdots     &  \vdots    & \vdots & \ddots \\
\end{array}%
\right).
\]

Obviously $\rho$ is computable, so it is lower-semicomputable. On the other hand, since $\rho$ is rank-one projection, if there is $\sigma$ such that $0 \leq \sigma \leq \rho$ then $\sigma = c \rho$ $(0 \leq c \leq 1)$. But it holds that $\braket{\lambda|\rho\lambda} \notin \mathbb{C}_q$ or $\braket{0|\rho\lambda} \notin \mathbb{C}_q$ for any $c \in \mathbb{R} \backslash 0$.

The same thing happens even if we allow a lower-approximation function to take an algebraic number, as G\'acs proposed in [3]. The operator

\[
\frac{1}{1 + \pi^2}
\left(%
\begin{array}{cccc}
  1            & \pi          & 0        & \ldots \\
  \pi          & \pi^2       & 0        & \ldots \\
  0            & 0            & 0        & \ldots \\
  \vdots     &  \vdots    & \vdots & \ddots \\
\end{array}%
\right)
\]
cannot be approximated by any sequence of positive operators from below. \qed
\end{ex}

%%%%%%%%%%%%%%%3.2%%%%%%%%%%%%%%%%%%
\subsection{Problem: the existence of a universal operator}

Just like the classical case, we expect that there is a universal semi-density operator in the following sense.

\begin{Def}
A lower-semicomputable semi-density operator $\mu$ is {\rm universal} if for any lower-semicomputable semi-density operator $\nu$ there is a real number $c_\nu>0$ such that $c_\nu \nu \leq \mu$.
\end{Def}

Unfortunately, our proof of the existence of a universal operator has somewhat weak form: namely, we need to assume some additional properties for each lower-approximation. We expect that these properties is derived from our definition.

Before stating the assumption and the proof, let us see the reason why we need such an additional assumption. In G\'acs [3], the following question is said to be solved positively in the same manner as the classical case, but it is not true.

\begin{prob}

Can we enumerate all lower-semicomputable semi-density operators effectively?

\end{prob}

To see the difficulty of this problem, let us review a proof of theorem 2.1.

\begin{prf11}

Let $\set{\varphi_n}_{n=1}^\infty$ be an effective enumeration of all partial recursive function. Consider the following algorithm: \\

{\bf Input} $n \in \mathbb{N}$.

\begin{enumerate}
  \item Let $\alpha_w \coloneqq 0$ for every $w \in \set{0,1}^*$.
  \item Dovetail $\varphi_n$, regarding  $\varphi_n$ as a function from $\set{0,1}^* \times \mathbb{N}$ to $\mathbb{Q}$. %英語表現
Whenever $\varphi_n$ halts for an input $\braket{w,k}$, go to step 3.
  \item Check whether the conditions $\varphi_n(w,k) \geq \alpha_w$ and $(\sum_{v \neq w} \alpha_v) + \varphi_n(w,k) \leq 1$ hold. If so, then let $\alpha_w \coloneqq \varphi_n(w,k)$. Otherwise, do nothing. go back to step 2.
\end{enumerate}

Let $\tilde \psi(w,t,n)$ be the value of $\alpha_w$ after the $t$-steps computation of the algorithm above for an input $n$. Obviously $\tilde \psi(-,-,n)$ is an lower-approximation of some lower-semicomputable semimeasure. $\tilde \psi$ can approximate any lower-semiconputable semimeasure from below, since any lower-approximation of a semimeasure is equal to some $\varphi_n$, and $\tilde \psi(-,-,n)$ approximates the same semimeasure from below. \qed

\end{prf11}

When we naively interpret this proof into the quantum setting, the corresponding algorithm would be like this:

{\bf Input} $n \in \mathbb{N}$.

\begin{enumerate}
  \item Let $\alpha_{w,v} \coloneqq 0$ for every $w,v \in \set{0,1}^*$, and let $\rho$ be an operator defined by $\braket{w|\rho v} \coloneqq \alpha_{w,v}$.
  \item Dovetail $\varphi_n$,  regarding  $\varphi_n$ as a function from $\set{0,1}^* \times\set{0,1}^* \times \mathbb{N}$ to $\mathbb{C}_q$. %英語表現
Whenever $\varphi_n$ halts for an input $\braket{w',v',k}$, go to step 3.

 \item Let $\rho '$ be an operator defined by
\[
\braket{w|\rho' v} \coloneqq
　\begin{cases}
　　\varphi_n(w,v,k) & ((w,v) = (w',v')) \\
　　\overline{\varphi_n(w,v,k)} & ((w,v) = (v',w')) \\
　　\alpha_{w,v} & (otherwise).
　\end{cases}
\] 
Check whether the condition $\rho' \geq \rho$ and ${\rm Tr}\rho' \leq 1$ holds. If so, then let $\alpha_{w',v'} \coloneqq \varphi_n(w',v',k)$ and $\alpha_{v',w'} \coloneqq \overline{\varphi_n(w',v',k)}$. Otherwise, do nothing. go back to step 2.
\end{enumerate}

For $\rho \in \mathbb{M}^n(\mathbb{C}_q)$ it is always possible to decide whether $\rho \geq 0$ or not (see [16]), so step 3 always ends in finite time. Let $\tilde \psi (w,v,t,n)$ be the value of $\alpha_{w,v}$ after the $t$-steps computation of the algorithm above for an input $n$.  The problem is that $\tilde \psi(-,-,-,n)$ generally does not approximate the same semi-density operator as which is approximated by $\varphi_n$ from below.

There are at least two main difficulties to construct the algorithm. First, updating process easily fails to maintain the monotonicity of sequence of operators, as long as we try to change the coefficients of the matrix pointwisely. For example, let $\varphi_{n_0}:\set{0,1}^* \times \set{0,1}^* \times \mathbb{N} \to \mathbb{C}_q$ be a recursive function such that

\[
\varphi_{n_0}(v,w,k)=
　\begin{cases}
　　\frac{1}{2} & (w,v \in \set{\lambda,0}) \\
　  0 & (otherwise),
　\end{cases}
\] 
and $t(\lambda, \lambda, 0) \ll t(1,1,0) \ll t(\lambda, 1, 0) \ll t$(any other input), where $t(v,w,k)$ is the time needed to compute $\varphi_{n_0}(v,w,k)$. We would expect that $\rho$ is updated as follows when we run the algorithm above for an input $n_0$, but it is not true:

\[
0 \to
\frac{1}{2}
\left(%
\begin{array}{cccc}
  1            & 0            & 0        & \ldots \\
  0            & 0            & 0        & \ldots \\
  0            & 0            & 0        & \ldots \\
  \vdots     &  \vdots    & \vdots & \ddots \\
\end{array}%
\right)
\to
\frac{1}{2}
\left(%
\begin{array}{cccc}
  1            & 0            & 0        & \ldots \\
  0            & 1            & 0        & \ldots \\
  0            & 0            & 0        & \ldots \\
  \vdots     &  \vdots    & \vdots & \ddots \\
\end{array}%
\right)
\to
\frac{1}{2}
\left(%
\begin{array}{cccc}
  1            & 1            & 0        & \ldots \\
  1            & 1            & 0        & \ldots \\
  0            & 0            & 0        & \ldots \\
  \vdots     &  \vdots    & \vdots & \ddots \\
\end{array}%
\right)
\]
Actually $\rho$ is never updated from the third step. Moreover, it turns out that for any $n \in \mathbb{N}$ the operator corresponds to $\lim_{k \to \infty}\tilde \psi(-,-,k,n)$ is diagonal. Hence, if we use the algorithm above, the expected-to-be-universal operator constructed in the same manner as the classical case is also diagonal, which cannot be universal (see proposition 3.13).

Second, as long as we initially set $\alpha_{w,v} \coloneqq 0$ for every $w,v \in \set{0,1}^*$, $\tilde \psi$ cannot be a lower-approximation of the operator described in the Example 3.5, since any $\tilde \psi (-,-,-,n)$ corresponds to a sequence of positive operators. 

%ここに改行をいれたい

To avoid these problems, we assume some additional properties for each lower-approximation. This is an analogous approach to the one of Tadaki [15], in which the notion of {\it lower-computable semi-POVM} is introduced, and it is shown that a universal semi-POVM does exist. The properties are as follows:

\begin{enumerate}
\item For a lower-approximation $\set{\rho_n}_{n=1}^\infty$ of any lower-semicomputable operator, each $\rho_n$ has a ``finite matrix representation with respect to the computational basis'': that is, there is a recursive function $f:\mathbb{N} \to \set{0,1}^*$ such that $P_{f(n)}\rho_nP_{f(n)} = \rho_n$, where $P_w \coloneqq \sum_{v=\lambda}^w \ket{v}\bra{v}$.
this property enables us to encode each $\rho_n$ to some natural number, and hence to avoid the difficulty to update the coefficients of the matrix pointwisely.

\item $\set{\rho_n}_{n=1}^\infty$ is a positive but ``almost increasing'' sequence: that is, there exists a computable density operator $\sigma$ such that for every $n \in \mathbb{N}$ it satisfies the conditions $\rho_n \geq 0$  and $\rho_{n+1} -\rho_n \geq -\rho^{-(n+1)}$. This is more restrictive than our definition since a sequence $\set{\rho_n - \sigma^{-n}}$ is always increasing and approximates the same element. 

It turns out that an operator $\frac{1}{2}(\rho + \sigma)$, which multiplicatively dominates $\rho$, is also lower-semicomputable semi-density and approximated by a sequence of positive operators. 
This property enables us to overcome an inability to find $n \in \mathbb{N}$ which makes $\tilde \psi$ a lower-approximation of certain operator.
\end{enumerate}

Here we restate our assumption more formally. We would like to call it conjecture since these properties are expected to be derived from our definition.

\begin{conj}
For given lower-semicomputable semi-density operator $\rho$, there exists a sequence  $\set{\rho_n}_{n=1}^\infty$ of operator which satisfies the following conditions:
\begin{enumerate}
\item $\rho_n \geq 0$, $\rho_n  \xrightarrow{n \to \infty} \rho$ (WOT), and there is a density operator $\sigma$ such that $\rho_{n+1} - \rho_n \geq -2^{-(n+1)}\sigma$.
\item There is a recursive function $\psi$ and $\varphi$ such that  $\psi(w,v,n) = \braket{w|\rho_nv}$  and $\varphi(w,v) = \braket{w|\sigma v}$.
\item There is a recursive function $f:\mathbb{N} \to \set{0,1}^*$ such that $P_{f(n)}\rho_nP_{f(n)} = \rho_n$.
\item For $\sigma_n \coloneqq P_{f(n)}\sigma P_{f(n)}$, it holds that $\sigma_{n+1}\geq \sigma_n$.
\end{enumerate}
\end{conj}

\begin{prop}
Assume the conjecture above is true. Then there exists a universal operator.
\end{prop}

\begin{prf}
First, we show an easy, but crucial fact.

\begin{cl}
Let $\rho \in \mathcal{\tilde S(H)}$ be lower-semicomputable, and $\set{\rho_n}_{n=1}^\infty$, $\sigma$, and $f$ be operators and a function described in conjecture 3.8, respectively. Then an operator $\rho' \coloneqq \frac{1}{2}(\rho + \sigma)$ is lower-semicomputable semi-density, and there exists a lower-approximation $\set{\rho'_n}_{n=1}^\infty$ of $\rho'$ which satisfies the conditions $\rho'_n \geq 0$ and $P_{f(n)}\rho'_nP_{f(n)} = \rho'_n$.
\end{cl} 

In fact, let $\rho'_n \coloneqq  \frac{1}{2}(\rho_n +(1-2^{-n})\sigma_n)$. Then the conditions $\rho'_n \geq 0$ and $P_{f(n)}\rho'_nP_{f(n)} = \rho'_n$ obviously hold, and showing $\rho'_n \xrightarrow{n \to \infty} \rho'$ is also straightforward. $\set{\rho'_n}_{n=1}^\infty$ is increasing since from the condition 1 and 3 of the conjecture we get
\begin{eqnarray*}
\rho_{n+1} + (1 - 2^{-(n+1)})\sigma_{n+1} 
	& = & P_{f(n+1)}(\rho_{n+1} + (1 - 2^{-(n+1)})\sigma)P_{f(n+1)} \\
	& \geq &  P_{f(n+1)}(\rho_n + (1 - 2^{-n})\sigma)P_{f(n+1)} \\
	& = & \rho_n + (1 - 2^{-n})\sigma_{n+1},
\end{eqnarray*}
and using the condition 4 of the conjecture we get $\rho'_{n+1} \geq \rho'_n$.

Now consider the following algorithm. Here, we let $\mathcal{L}_q(\mathbb{C}^m)$ be the set of all $m \times m$ hermitian matrices of which each coefficient is in $\mathbb{C}_q$, and often identify an operator in $\mathcal{L}_q(\mathbb{C}^m)$ with that on $\mathcal{H}$ in a canonical way.

{\bf Input} $n \in \mathbb{N}$.

\begin{enumerate}
  \item Let $\nu \coloneqq 0$  ($\nu \in \mathcal{B(H)}$).
  \item Dovetail $\varphi_n$, regarding  $\varphi_n$ as a function from $\mathbb{N}$ to $\bigcup_{m \in \mathbb{N}}\mathcal{L}_q(\mathbb{C}^m)$. %英語表現
Whenever $\varphi_n$ halts for an input $k$, go to step 3.
  \item Check whether the conditions $\varphi_n(k) \geq \nu$ and Tr$\varphi_n(k) \leq 1$ hold. If so, then let $\nu \coloneqq \varphi_n(k)$. Otherwise, do nothing. go back to step 2.
\end{enumerate}

Let $\tilde \psi(n,t)$ be the value of $\nu$ after the t-steps computation of the algorithm above for an input $n$. 
%Using lemma (hoge), 
It can be shown that for every $n \in \mathbb{N}$ there exists $\nu_n \in \mathcal{B(H)}$ such that  $\tilde \psi(n,t) \xrightarrow{t \to \infty} \nu_n$ in WOT.
Obviously $\nu_n$ is lower-semicomputable semi-density.
%For any $n \in \mathbb{N}$, $\tilde \psi(n,t)$ converges to a semi-density operator $\nu_n$ as $t \xrightarrow{} \infty$  (notice that we use lemma (hoge) to show the sequence $\set{\tilde \psi(n,t)}_{n=1}^\infty$ converges in WOT). 

Now let $\rho \in \mathcal{\tilde S(H)}$ be lower-semicomputable, and $\set{\rho_t}$, $\sigma$, and $f$ be operators and a function described in conjecture 3.8, respectively.
Then there is $n \in \mathbb{N}$ such that $\nu_n = \frac{1}{2}(\rho + \sigma)$. 
In fact, there exists $n \in \mathbb{N}$ such that $\varphi_n(t) = \rho'_t$, where $\rho'_t$ is described in the claim above, and $\set{\tilde \psi(n,t)}_{t=1}^\infty$ is also a lower-approximation of $\rho'$. This can be shown using the fact that $\set{\tilde \psi(n,t)}_{t=1}^\infty = \set{\rho'_{g(n)}(t)}_{t=1}^\infty$, where $g : \mathbb{N} \to \mathbb{N}$ is an appropriate nondecreasing, unbounded function (We assume $\rho'_1 = 0$ without loss of generality). 

Finally, we can show $\mu \coloneqq \sum_{n=1}^\infty 2^{-n}\nu_n$ is universal in the following way:
\begin{itemize}
\item Since $\set{ \sum_{k=1}^n 2^{-k}\nu_k}_{n=1}^\infty$ is a Cauchy sequence, $\mu$ is well-defined semi-density operator. 

\item $\mu$ dominates any lower-semicomputable semi-density operator $\rho$, since there is $n \in \mathbb{N}$ such that $2\nu_n = \rho + \sigma$, so $\rho \leq 2\nu_n \leq 2^{(n+1)}\mu$.

\item $\mu$ is also lower-semicomputable since $\varphi (n,w,v) \coloneqq \sum_{k=1}^n 2^{-k}\psi_k(n,w,v)$ is its lower-approximation, where $\psi_k$ is a lower-approximation of $\nu_k$.
In fact, for given $\epsilon > 0$ and a unit vector $\ket{\psi} \in \mathcal{H}$, there is an integer $k_0$ such that $\| \sum_{k=k_0+1}^\infty 2^{-k} \nu_k \| < \epsilon$, and there is an integer $k_1 \geq k_0$ such that $\braket{\psi|(\nu_n - \nu_{nk_1})\psi} < \frac{2^n}{k_0} \epsilon$ for every $n \leq k_0$. Hence
\begin{eqnarray*}
\braket{\psi |(\mu - \mu_{k_1}) \psi}
&\leq& \braket{\psi | \mu \psi} - \sum_{n=1}^{k_0}\braket{\psi | \nu_{nk_1} \psi}\\
& = & \sum_{n=1}^{k_0} 2^{-n} \braket{\psi | (\nu_n - \nu_{nk_1})\psi} + \braket {\psi | (\sum_{n=k_0 + 1}^\infty 2^{-n} \nu_n) \psi}\\
& < & 2\epsilon ,
\end{eqnarray*}
so $\mu_n \xrightarrow{} \mu$ in WOT. Obviously $\mu_n$ is increasing, and $\varphi(n,w,v) \in \mathbb{C}_q$ for every $n \in \mathbb{N}$ and $w,v \in \set{0,1}^*$. \qed
\end{itemize}
\end{prf}

%%%%%%%%%%%%%%%%%%%%%%%%%%

Once we prove the existence of universal semi-density operator, we can define quantum algorithmic entropy $\overline{H}$ and $\underline{H}$ in the same manner as G\'acs [3]:
\[
 \overline{H}(\ket{\psi}) = -\braket{\psi|(\log \mu)\psi}, \mbox{    }  \underline{H}(\ket{\psi}) =-\log \braket{\psi|\mu\psi}.
\]

The following proposition claims that  $\overline{H}$ and  $\underline{H}$ are the extensions of classical descriptive complexity. 

\begin{prop}
Assume a universal operator $\mu$ exists. Then for any uniformly computable orthonormal system $\set{\ket{\psi_n}}_{n=1}^\infty$  (not necessarily a basis), 
\[
K(n) = H(\ket{\psi_n}) + O(1),
\]
where $H=\overline{H}$ or $H=\underline{H}$. In particular, for any $w \in \set{0,1}^*$,
\[
K(w) = H(\ket{w}) + O(1).
\]
\end{prop}

We strongly expect this equation holds, since there is an analogous consequence about qubit complexity defined by Berthiaume et al [2]. Our eventual goal is to examine the equivalence of qubit complexity and G\'acs' quantum algorithmic entropy, so this is a very minimum requirement for us.

\begin{prop}[{[8]}]
For any $w \in \set{0,1}^*$,
\[
C(w) = QC(\ket{w}) + O(1).
\]
Here, $QC(\ket{\psi})$ is the qubit complexity of $\ket{\psi}$ (see {\rm [2]}).
\end{prop}

\begin{prf46}
The proof is completely the same as the one in [3], but it is valid in our definition. The function $f(n) = \braket{\psi_n|\mu \psi_n}$ is lower-semicomputable with $\sum_n f(n) = \mbox{\rm Tr } \mu \leq 1$, hence $K(n) \leq \underline{H}(n) + O(1)$.

On the other hand, the semi-density operator $\rho = \sum_n {\rm \bf m}(n) \ket{\psi_n} \bra{\psi_n}$ is lower-semicomputable (proposition 3.4), so

\[
K(n) = \braket{\psi_n|(-\log \rho)\psi_n} \geq \braket{\psi_n|(-\log \mu)\psi_n} + O(1) = \overline{H}(\ket{\psi_n}) + O(1).
\]
Notice that the inequality above holds since $g(x) = \log x $ is an operator monotone function. Finally, for any state $\ket{\psi}$ we have an inequality $\overline{H}(\ket{\psi_n}) \geq \underline{H}(\ket{\psi_n})$, which completes the proof.
 \qed
\end{prf46}

\begin{rem}
The statement which makes the problem in the definition by G\'acs is `` $\sum_n f(n) = \mbox{\rm Tr } \mu \leq 1$''. His universal operator is actually the sequence $\set{\mu_n}_{n=1}^\infty$ of matrices, and $\mu$ in the definiton of $f$ is actually some appropriate $\mu_{k_n}$. The value of $k_n$ cannot be the same for all $n \in \mathbb{N}$ since $\set{\ket{\psi_n}}_{n=1}^\infty$ is an infinite sequence of orthogonal states. So we do not know how to show $\sum_n f(n) \leq 1$. Also we do not know what the statement ``$\rho = \sum_n {\rm \bf m}(n) \ket{\psi_n} \bra{\psi_n}$ is lower-semicomputable'' means in his finite dimensional formulation. In short, the proof is stated as if we work on an infinite dimensional setting, and it is one of the main reasons we try to modify his definition into an infinite dimensional version.
\end{rem}

We conclude this subsection with an easy corollary which evokes an analogous fact in classical domain: for a universal semimeasure ${\bf m}$ and an infinite recursive set $\set{w_n} \subset \set{0,1}^*$, a function ${\bf m}'(n) \coloneqq {\bf m}(w_n)$ is again universal. The following seems to be the quantum version of this fact.

\begin{cor}
Assume a universal operator $\mu$ exists. Let $\set{\ket{\psi_n}}$ be a uniformly computable orthonormal system of $\mathcal{H}$. Then a function ${\bf m}_{\psi}(n) \coloneqq \braket{\psi_n|\mu \psi_n}$ is a universal semimeasure.
\end{cor} 

%%%%%%%%%%%%%%3.3%%%%%%%%%%%%%%

\subsection{$\mu$ is not diagonal}

At first glance, one might expect that an operator $\mu_1 \coloneqq \sum_i {\bf m}(i)\ket{i}\bra{i}$ is universal. In fact, for corollary 3.12, diagonal entries of universal operator should form a universal semimeasure, so it would be natural to question whether the simplest operator with this property, i.e. a diagonal one, is universal.

It is rather disappointing if the answer is yes, since in this case $H$ is very simple combination of classical complexity: 

\[
\overline H(\sum_w \alpha_w \ket{w}) = - \sum_w  |\alpha_w|^2 {\rm log}{\bf m}(w) \mbox{ , } \underline H (\sum_w \alpha_w \ket{w}) = - {\rm log}\sum_w  |\alpha_w|^2 {\bf m}(w).
\]
For good or bad, it turns out $\mu_1$ is not universal.

\begin{prop}
There is a lower-semicomputable semi-density operator which cannot be multiplicatively dominated by $\mu_1$.
\end{prop}

\begin{prf}
Assume $\mu_1$ is universal, and let  $\ket{\psi_n} \coloneqq 2^{-\frac{n}{2}} \sum_{l(w)=n} \ket{w}$. Then for corollary 3.12, the function $\overline{\bf m}(n) \coloneqq \braket{\psi_n|\mu_1 \psi_n} = 2^{-n}\sum_{l(w)=n}  {\bf m}(w)$ must be a universal semimeasure, which is not true. In fact, The function $2^n\overline{\bf m}$ is also a lower-semicomputable semimeasure which cannot be dominated by $\overline{\bf m}$. \qed
\end{prf}

%This consequence evokes an analogious fact in classical domain: for a universal semimeasure ${\bf m}$ and an infinite recursive set $\set{w_n} \subset \set{0,1}^*$, a function ${\bf m}'(n) \coloneqq {\bf m}(w_n)$ is again universal. We could explain the reason why $\rho$ in proposition 4.9 works as a counterexample is that a semimeasure $m(n) = 2^n \overline{\bf m}(n)$ is not universal. The counterexample tells us that we can "mine" more various semimeasures from lower-semicomputable semi-density oprator compared with lower-semicomputable semimeasure itself, so $\mu$ should have very complicated form in some sense, if it exists. 

We can derive a more general fact which tells us the set of eigenspaces and eigenvalues of $\mu$ should have certain "incomputablilty".

\begin{cor}
There is no uniformly computable orthonormal basis $\set{\ket{\psi_n}}_{n=1}^\infty$ of $\mathcal{H}$ and lower-semicomputable semimeasure $m$ which makes an operator $\sum_n m(n){\ket{\psi_n}}{\bra{\psi_n}}$ universal.
\end{cor}

\begin{prf}
Let $\ket{\varphi_n} \coloneqq 2^{-\frac{n}{2}} \sum_{l=2^(n-1)}^{2^n-1} \ket{\psi_l}$ and consider the same argument as the previous proof. \qed
\end{prf}
%%%%%%%%%%%%%%%%%%%%%%%%%%%%

%\input{./sec4.tex}

\section{Discussion and perspective}

We defined an infinite dimensional modification of lower-semicomputability of density operators by G\'acs, and examined their properties, especially the well-definedness of his quantum algorithmic entropy. We needed some additional assumption to establish well-defined notion, and checking whether this assumption can be eliminated or not is left as a future task. 

In particular, the condition 1 of conjecture 3.8 could be relaxed or eliminated in some way. As we saw in the proof of proposition 3.9, for given $\rho$, we only needed to find $\nu_n$ which multiplicatively dominates $\rho$, not which is equal to $\rho$ itself. The necessity of the condition 1 has arisen from example 3.5, but this operator is actually dominated by $\ket{\lambda} \bra{\lambda} + \ket{0} \bra{0}$, which is apparently positively-approximated lower-semicomputable operator. It is likely that there is some nice algorithm to find a dominating, positively-approximated operator, for given $\rho$.

%Eventually, we would like to examine whether the Levin's coding theorem holds in quantum domain. To do this, we also need to define a “prefix version” of qubit complexity by Berthiaume et al. Proposition 3.11 says that qubit complexity is an extension of plain Kolmogorov complexity, not prefix-free one, while Levin's theorem indicates equivalence of prefix-free Kolmogorov complexity and universal semimeasure.

%\input{./tadaki.tex}
\section*{Acknowledgment}
The author would like to thank 
his supervisor Prof. Masahito Hasegawa for constant support and encouragement,
Prof. Kohtaro Tadaki, Prof. Peter G\'acs and Prof. Takayuki Miyadera for helpful discussions, and
two anonymous referees for the valuable comments.

\providecommand{\urlalt}[2]{\href{#1}{#2}} \providecommand{\doi}[1]{doi:\urlalt{http://dx.doi.org/#1}{#1}}

%\nocite{*}
%\bibliographystyle{eptcs}
%\bibliography{generic}
\end{document}